\documentclass[12pt]{article}
\usepackage[dvips]{graphicx}
\usepackage{amssymb}
\usepackage{amsmath}
\usepackage{epsfig}
\usepackage{cite}

\numberwithin{equation}{section}
\numberwithin{table}{section}

\def\beq{\begin{equation}}
\def\eeq{\end{equation}}
\def\be{\begin{equation}}
\def\ee{\end{equation}}
\def\bea{\begin{eqnarray}}
\def\eea{\end{eqnarray}}

\def\SO{\textrm{SO}}
\def\SU{\textrm{SU}}
\DeclareRobustCommand{\SkipTocEntry}[4]{}
\RequirePackage{color}

\newcommand{\cO}{\mathcal{O}}

\newcommand{\cL}{\mathcal{L}}

\newcommand{\cM}{\mathcal{M}}
\newcommand{\cN}{\mathcal{N}}

\newcommand{\bbN}{\mathbb{N}}
\newcommand{\bbZ}{\mathbb{Z}}
\newcommand{\bbR}{\mathbb{R}}

\newcommand{\dd}{\mathrm{d}}
\newcommand{\D}{\mathrm{D}}     
\def\AdS{\textrm{AdS}}

\begin{document}
\begin{titlepage}
\begin{center}
\rightline{\small ZMP-HH/15-17}

\vskip 1cm

{\Large \bf Supersymmetric AdS$_7$ backgrounds in 
half-maximal supergravity 
and marginal operators of $(1,0)$ SCFTs
}
\vskip 1.2cm

{\bf  Jan Louis 
and Severin L\"ust }

\vskip 0.8cm

{\em Fachbereich Physik der Universit\"at Hamburg, Luruper Chaussee 149, 22761 Hamburg, Germany}
\vskip 0.2cm

and 

\vskip 0.2cm

{\em Zentrum f\"ur Mathematische Physik,
Universit\"at Hamburg,\\
Bundesstrasse 55, D-20146 Hamburg, Germany}
\vskip 0.3cm

\vskip 0.3cm

{\tt jan.louis@desy.de, severin.luest@desy.de}

\end{center}

\vskip 1cm

\begin{center} {\bf ABSTRACT } \end{center}

\noindent
We determine the supersymmetric AdS$_7$ backgrounds of
seven-dimensional half-maximal gauged supergravities
and show that they do not admit any deformations that preserve all 16
supercharges. 
We compare this result to the conformal manifold of the holographically dual $(1,0)$ superconformal field theories 
and show that accordingly its representation theory implies that no
supersymmetric  marginal operators exist.

\vfill


June 2015

\end{titlepage}


\tableofcontents


\section{Introduction}

AdS backgrounds of gauged supergravities have been prominently studied 
in connection with the AdS/CFT correspondence
\cite{Maldacena:1997re}. 
In particular a large variety of explicit solutions  
of ten- and eleven-dimensional supergravities of the form 
$\AdS_d\times Y_{10/11-d}$ have been constructed by now. 
Generalization of the original $AdS_5\times S^5$ started in 
refs.~\cite{Kehagias:1998gn,Morrison:1998cs} and the more recent developments are
summarized, for example, in \cite{Polchinski:2010hw}.
Refs.~\cite{deAlwis:2013jaa,Louis:2014gxa} on the other hand studied $\AdS_4$
backgrounds within
four-dimensional ($d=4$) supergravities without considering any explicit
relation with solutions of higher-dimensional supergravities.
It was found that the existence of AdS backgrounds imposes specific
conditions on the couplings of the supergravity.
For $\cN=1$ these conditions are formulated in terms of the K\"ahler
potential and the superpotential. For $\cN>1$ AdS backgrounds can only
appear in gauged supergravities and the necessary gaugings are
conveniently expressed in terms of the embedding tensor 
\cite{Nicolai:2000sc,deWit:2002vt}. 
Concretely refs.~\cite{deAlwis:2013jaa,Louis:2014gxa} studied  $\cN=2$ and
$\cN=4$ AdS backgrounds together with their  
deformations that preserve all
supercharges  and determined the structure and properties of this
moduli space.
For $\cN=4$ it is even possible to classify the $\AdS_4$ backgrounds in
that the structure of possible gauge groups can be given. 
In particular a specific subgroup of the R-symmetry group always has to be gauged and has to be unbroken in the AdS background.
Furthermore it was shown that
no deformations preserving all 16 supercharges exist and one can only
have isolated vacua. In \cite{Louis:2015dca} the analysis  was carried over
to $\AdS_5$ backgrounds of five-dimensional gauged supergravities
with 16 supercharges where a similar classification is possible but in
this case a moduli space does exist.

In this paper we extend these studies to seven-dimensional supergravities with
sixteen supercharges (half-maximal) coupled to an arbitrary number of
vector multiplets and determine their $\AdS_7$ backgrounds. 
Unfortunately, the most general Lagrangian 
of these supergravities
formulated in the embedding tensor formalism is not yet know. 
The original papers
\cite{Townsend:1983kk,Bergshoeff:1985mr} take some 
of the embedding tensor components into account but not all.
This has been partly remedied in
\cite{Bergshoeff:2005pq,Bergshoeff:2007vb,Dibitetto:2015bia} where
all embedding tensor components have been identified. However, so far
only certain terms of the  full Lagrangian for these additional 
embedding tensor components have been given.
Luckily we can show that in supersymmetric $\AdS_7$ backgrounds
only specific embedding tensor components can be non-trivial and for these
the Lagrangian is known.

While we work solely within the framework of seven-dimensional gauged supergravity, supersymmetric $\AdS_7$ solutions can be also discussed from the perspective of higher-dimensional supergravity.
The only half-maximally supersymmetric solutions in M-theory are of the form \(AdS_7 \times S^4/\bbZ_k\) \cite{Ferrara:1998vf, Ahn:1998pb}. There are no supersymmetric $AdS_7$ solutions in type IIB supergravity, but solutions of massive type IIA supergravity have been explicitly constructed and classified in \cite{Apruzzi:2013yva,Apruzzi:2015wna,Apruzzi:2015zna}.
All these solutions can be truncated consistently to minimal gauged supergravity in seven dimensions \cite{Passias:2015gya} and should hence describe a possible higher-dimensional origin for the solutions discussed in this paper.\footnote{In \cite{Danielsson:2013qfa}
it was however noted that certain solutions of ten-dimensional type IIA string
theory (including localized and smeared branes) do not seem to have a
description within seven-dimensional gauged supergravity.}

In our analysis we find that supersymmetric $\AdS_7$ backgrounds require
the gauge group $G$ to be of the general form 
\begin{equation}\label{result}
G\ =\ G_0\times H\subset SO(3,n)\ ,
\end{equation}
 where $n$ is the number of vector
multiplets, $H$ is a compact semi-simple factor while 
$G_0$ needs to contain an $SO(3)$ subgroup which has to coincide with the unbroken gauge group in the vacuum.
The unbroken $SO(3)$ is the 
R-symmetry of the supergravity or an admixture of the R-symmetry 
with an appropriate $SO(3)$ factor associated with the vector
multiplets.\footnote{Contrary to $\AdS_4$ and $\AdS_5$ backgrounds
with 16 supercharges, in $d=7$ the entire R-symmetry group $SO(3)$ has to be gauged and unbroken.}
If we assume the gauge group to be semi-simple, we can further restrict \(G_0\) to be either $SO(3)$, $SO(3,1)$ or
$SL(3, \bbR)$.
A related result has been obtained in \cite{Karndumri:2014hma,
  Karndumri:2014sha} from a different approach.

Furthermore, we study the scalar deformations of the
AdS backgrounds which
preserve all supercharges and show that they all are Goldstone bosons
of the 
spontaneously broken gauge group $G$ and therefore do not count as
physical moduli.
Consequently there is no supersymmetric moduli space exactly as for
$d=4$, $\cN=4$ \cite{Louis:2014gxa}.

In the second part of the  paper
we consider the holographically dual six-dimen\-sional $\cN=(1,0)$
superconformal field theories (SCFT) and their possible exactly
marginal deformations. This deformation space is known as  the
conformal manifold and according to the AdS/CFT correspondence it
should coincide with the moduli space of the  AdS solutions. 
In agreement with our previous results we can indeed show 
that there is no \(\cN = (1,0)\) SCFT that can have any
supersymmetric marginal deformations and thus no conformal manifold exists.
This follows solely from the representation theory of the
superconformal algebra in that any possible marginal operator violates
the unitarity bounds and therefore is forbidden.
Here we essentially follow a similar analysis for $\cN=1$ SCFTs in $d=4$
performed in \cite{Green:2010da} and 
use
 the \(\cN = (1,0)\) representation theory determined in
\cite{Minwalla:1997ka,Dobrev:2002dt,Bhattacharya:2008zy}.
When this manuscript was being completed we learned about
ref.~\cite{Intriligatoretal} which has considerable overlap with the
second part of this paper.

This paper is organized as follows. In section~\ref{sugraintro} we
briefly review the half-maximal supergravities in $D=7$.
In \ref{sec:adsbackgrounds} we 
show that supersymmetric $\AdS_7$
backgrounds imply the gauge group given in \eqref{result}.
In \ref{sugramoduli} we show that the resulting scalar potential has
flat directions but all of them correspond to Goldstone bosons of a
spontaneously broken gauge group in the $\AdS_7$ vacuum.
In section~\ref{sec:SCFT} we turn to the dual superconformal theories.
After discussing some general properties  in \ref{prelim}
we show in section~\ref{sec:marginaloperators} that 
there are no
marginal operators in six-dimensional $\cN = (1,0)$ SCFTs.
In Appendix~\ref{app:superconfalgebra} we review the six-dimensional $\cN = (1,0)$ superconformal algebra,
 and in Appendix~\ref{app:minlevel} we discuss the group theoretical restrictions on the level of a Lorentz invariant descendant operator.


\section{$\AdS_7$ backgrounds of seven-dimensional half-maximal
  supergravity} 
\label{sugra}

\subsection{Preliminaries}
\label{sugraintro}
In this section we briefly recall the structure of 
half-maximal gauged supergravities in \(d=7\) following
\cite{Bergshoeff:1985mr,Bergshoeff:2005pq,Bergshoeff:2007vb,Dibitetto:2015bia}.
The gravity multiplet has the field content
\begin{equation}
\left(g_{\mu\nu}, \psi^A, A^i_\mu, \chi^A, B_{\mu\nu}, \sigma\right)
\,,\qquad \mu,\nu=0,\ldots,6\ ,
\end{equation}
where $g_{\mu\nu}$ is the metric, \(\psi^A\), \(A=1,2\) is an $SU(2)_R$-doublet of
gravitini, \(A^i_\mu\), \(i=1,2,3\), is
an R-triplet of vectors, \(\chi^A\) is an $SU(2)_R$-doublet of spin-1/2 fermions, 
\(B_{\mu\nu}\) is an antisymmetric tensor  and \(\sigma\) a real scalar.
Furthermore there can be \(n\) vector multiplets
\begin{equation}
\left(A^r_\mu, \lambda^{rA}, \phi^{ri}\right) \,,\qquad r=1,\dots,n\ , 
\end{equation}
where each consists of one vector \(A^r_\mu\), a doublet of spin-1/2
fermions \(\lambda^{rA}\) and a triplet of real scalars \(\phi^{ri}\).
All fermions are symplectic Majorana spinors.
Altogether there are $(n+3)$ vector
fields,
$(2n+2)$ spin-1/2 fermions and $(3n+1)$ real scalars.

The field space $\cM$ of the 
scalars 
is given by 
\begin{equation}\label{eq:cosetspace}
\cM\ =\  \mathbb{R}^+ \times\frac{SO(3, n)}{SO(3) \times SO(n)}\,\,,
\end{equation}
where the $3n$ dimensional coset manifold is spanned by the scalars \(\phi^{ri}\)
in the vector multiplet while the $\mathbb{R}^+$ factor 
corresponds to  $\sigma$.
The coset can be conveniently parametrized by a coset
representative\footnote{In the recent literature on gauged
  supergravity the coset
  representatives are often denoted 
by ${\cal V}^i_I$. Here we choose the notation of the original papers
\cite{Bergshoeff:1985mr,Bergshoeff:2005pq,Bergshoeff:2007vb,Dibitetto:2015bia}.}
\begin{equation}
L = \left(L^i_I, L^r_I\right) \,,\quad I=1,\dots,n+3 \ .
\end{equation} 
$L$ is an \(SO(3,n)\) matrix and hence satisfies
\begin{equation}\label{eq:eta}
\eta_{IJ} = -L^i_I L^i_J + L^r_I L^r_J \,,
\end{equation}
where \(\eta_{IJ} = \mathrm{diag}(-1,-1,-1,+1,\dots,+1)\) is the
canonical \(SO(3,n)\) metric.
The scalar manifold $\cM$ can be described by the metric 
\begin{equation}
M_{IJ} = L^i_I L^i_J + L^r_I L^r_J \,.
\end{equation}

The 
\((n+3)\) vector fields are combined into
\(A^I = (A^i, A^r)\) and can be rotated into each other by the global symmetry group \(SO(3,n)\).
A subgroup \(G \subset SO(3,n)\) can be made local provided
that the structure constants \({f_{IJ}}^K\) of \(G\) are completely
antisymmetric, i.e.\ they satisfy the linear constraint
\begin{equation}\label{eq:gaugecondition}
{f_{IK}}^L\eta_{LJ} + {f_{JK}}^L\eta_{LI} = 0\ .
\end{equation}
Clearly the dimension of \(G\) is restricted by the number of vectors fields to be not larger than \(n+3\).
As explained in \cite{deRoo:1985jh} the condition \eqref{eq:gaugecondition} restricts the choice of possible (non-compact) gauge groups \(G\) strongly.
Since \(\eta_{IJ}\) has signature \((3,n)\) any semi-simple \(G\)
can be either generated by at most three compact or 
three
non-compact generators
and the allowed semi-simple gauge groups are cataloged in
\cite{Bergshoeff:2005pq}.
In  the next section we will determine which of the
gauge groups can give rise to AdS vacua. 

To construct a gauge invariant action it is convenient to introduce the gauged Maurer-Cartan one-forms
\begin{equation}\begin{aligned}\label{eq:maurercartan}
P^{ir} &= L^{Ir} (\delta_I^K \dd + {f_{IJ}}^K A^J) L^i_K \,, \\
\end{aligned}\end{equation}
where  $L^{Ir}$ denotes the inverse coset representative. 
The gauge covariant field strengths are defined by
\begin{equation}
F^I = \dd A^I + \frac{1}{2}{f_{JK}}^I A^J \wedge A^K \,.
\end{equation}
Furthermore, for the existence of AdS vacua it turns out to be necessary to dualize the two-form \(B_2\) into a
three-form \(G_3\) 
 and to add to the action
the topological mass term \cite{Bergshoeff:2005pq}
\begin{equation}
S_h = 4h \int H_4 \wedge G_3 \ ,
\end{equation}
where $h$ is a real constant
and \(H_4 = \dd G_3\) is the four-form field strength.

With these ingredients the total bosonic Lagrangian of gauged \(\cN = 2\) supergravity reads \cite{Bergshoeff:1985mr, Bergshoeff:2005pq}
\begin{equation}\begin{aligned}\label{eq:lagrangian}
\cL\ =\ &\frac{1}{2} R \ast\! 1- \frac{1}{2} e^\sigma M_{IJ} F^I \wedge \ast F^J - \frac{1}{2}e^{-2\sigma} H_4 \wedge \ast H_4 - \frac{5}{8} \dd \sigma \wedge \ast \dd \sigma \\
&- \frac{1}{2} P^{ir} \wedge \ast P_{ir} - \frac{1}{\sqrt2} H_4 \wedge \omega_3 + 4h\,  H_4 \wedge G_3 - V \ast\! 1 \,,
\end{aligned}\end{equation}
where the Chern-Simons three-form \(\omega_3\) is given by
\begin{equation}
\omega_3 = \eta_{IJ} F^I \wedge A^J - \frac{1}{6}{f_{IJ}}^K A^I \wedge
A^J \wedge A_K \ .
\end{equation}
The potential $V$ takes the form
\begin{equation}\label{eq:potential}
V = \frac{1}{4} e^{-\sigma} \left(C^{ir} C_{ir} - \frac{1}{9}
   C^2\right) + 16 h^2 e^{4\sigma} - \frac{4\sqrt2}{3}\,h\,
e^\frac{3\sigma}{2} C \ ,
\end{equation}
where we abbreviated
\begin{equation}\begin{aligned}\label{abbreviations}
C &= -\frac{1}{\sqrt2}\, f_{ijk}\epsilon^{ijk}
\,, \qquad C_{ir} = \frac{1}{\sqrt2}\, f_{jkr}\epsilon^{ijk} \,,
\\[1ex]
 f_{ijk}&={f_{IJ}}^{K} L^I_i L^J_j L_{Kk}\ ,\qquad
 f_{jkr}= {f_{IJ}}^{K} L^I_j L^J_k L_{Kr}\ .
\end{aligned}\end{equation}

Finally, to find the background solutions that preserve supersymmetry we need the supersymmetry variations of all fermionic fields. They are given by
\begin{equation}\begin{aligned}\label{eq:susyvariations}
\delta\psi_\mu &= \D_\mu \epsilon - \frac{\sqrt2}{30}\, e^{-\frac{\sigma}{2}} C \gamma_\mu \epsilon - \frac{4}{5}\, h e^{2\sigma} \gamma_\mu \epsilon + \dots \,, \\
\delta\chi &= \frac{\sqrt2}{30}\, e^{-\frac{\sigma}{2}} C \epsilon - \frac{16}{5}\, e^{2 \sigma} h \epsilon + \dots \,, \\
\delta\lambda^r &= - \frac{i}{\sqrt2}\, e^{-\frac{\sigma}{2}} C^{ir} \sigma^i \epsilon + \dots \,,
\end{aligned}\end{equation}
where we suppressed the R-symmetry index \(A\) and the ellipses
denote terms which vanish in a maximally symmetric
space-time background.

So far we used the supergravity as determined in
\cite{Bergshoeff:1985mr,Bergshoeff:2005pq}. However,
ref.~\cite{Bergshoeff:2007vb} pointed out that this is not the most
general formulation of gauged \(\cN=2\) supergravity because apart
from the totally antisymmetric structure constant \(f_{[IJK]}\)  
there can also be another gauge parameter \(\xi_I\)
which  transforms in the vector representation of \(SO(3,n)\).
Denoting the generators of \(SO(3,n)\) by \(t_{[IJ]}\) and the
generator of the \(\bbR^+\) shift symmetry of \(\sigma\) by \(t_0\),
the full embedding tensor is then given by \cite{Bergshoeff:2007vb}
\begin{equation}\begin{aligned}
{\Theta_I}^{JK} &= {f_I}^{JK} + \delta_I^{[J}\xi^{K]} \ , \qquad\quad
{\Theta_I}^0 &= \xi_I \ ,
\end{aligned}\end{equation}
and the general covariant derivative reads
\begin{equation}
\D = \dd - A^I {f_I}^{JK} t_{JK} - A^I \xi^J t_{IJ} - A^I \xi_I t_0 \ .
\end{equation}
With this information one can determine
the Lagrangian of the supergravity. A partial answer has been obtained 
recently in \cite{Dibitetto:2015bia} but the full  Lagrangian has not
been given yet. Luckily, we will see in the next section that
supersymmetric AdS solutions can only occur for \(\xi_I=0\), so that in fact we do not need to use the most general formulation.  
In order to show this we will need 
 the additional \(\xi_I\) dependent terms in the supersymmetry
 variations given in \eqref{eq:susyvariations}.
They are of the form \cite{Dibitetto:2015bia}
\begin{equation}\label{eq:susyvariationsxi}
\delta\chi \sim e^{-\frac{\sigma}{2}}\xi^i \sigma^i\epsilon + \ldots\,,\qquad
\delta\lambda^r \sim e^{-\frac{\sigma}{2}}\xi^r\epsilon  + \ldots\,,
\end{equation}
where $\xi^i= L^i_I\xi^I$, $\xi^r= L^r_I\xi^I$. These variations in turn
induce an additional term in the potential given by 
\begin{equation}
V_\xi \sim e^{-\sigma}\left(\xi^i\xi^i + \xi^r\xi^r\right) = e^{-\sigma}\xi_I \xi_J M^{IJ} \,.
\end{equation}

\subsection{Supersymmetric AdS backgrounds}\label{sec:adsbackgrounds}
In this section we derive conditions on the gauge group \(G\) such that the theory admits fully supersymmetric AdS vacua.
Unbroken supersymmetry implies that the supersymmetry variations of the
fermions \eqref{eq:susyvariations} and \eqref{eq:susyvariationsxi} vanish in the AdS background and
therefore we need to have  
\begin{equation}\label{eq:Cconditions}
\left<C_{ir}\right> = 0 \,,\qquad
\left<C\right> = \frac{96}{\sqrt2}\, h\, e^{\frac{5}{2}\left<\sigma\right>} \,,\qquad
\left<\xi^i\right> = \left<\xi^r\right> = 0 \, .
\end{equation}
As promised we find \(\left<\xi^I \right> =0\) 
which follows from the fact that $({\bf 1},\sigma^i)$ forms a  basis
of two-dimensional Hermitian matrices  and thus
the terms given in \eqref{eq:susyvariationsxi}
cannot cancel against terms in \eqref{eq:susyvariations}. 
Using the ``dressed'' structure constants  
defined in \eqref{abbreviations} the first two conditions in 
\eqref{eq:Cconditions}
read
\begin{equation}\begin{aligned}\label{eq:strconstcond}
\left<f_{ijk}\right> = -g \epsilon_{ijk}\,, \qquad
\left<f_{ijr}\right> = 0 \,,
\end{aligned}\end{equation}
where the coupling constant \(g\) can be chosen arbitrarily and dictates together with \(h\) the value of the cosmological constant.
Inserted into \eqref{eq:potential}, the cosmological constant is 
\begin{equation}
\Lambda = \left<V\right> = -240 h^2 e^{4\left<\sigma\right>}\ ,
\end{equation}
and we indeed
see that the background is AdS if and only if a topological mass term
with coupling $h$ is included into the action \cite{Karndumri:2014sha, Karndumri:2014pla}.
We also see from  \eqref{eq:Cconditions} and \eqref{eq:strconstcond}
that the scalar \(\sigma\) from the gravity multiplet has to take the
background value
\begin{equation}
\left<\sigma\right> = \frac{2}{5}\log{\left(\frac{g}{16h}\right)} \,.
\end{equation}

The conditions \eqref{eq:strconstcond} on the structure constants are
very similar to those derived in \cite{Louis:2014gxa} so that we can
essentially follow their 
analysis for determining the gauge group.
The simplest situation occurs when in addition to \eqref{eq:strconstcond}
there are no mixed index components of the structure constants, i.e.\
\(f_{ist} = 0\). In this case the gauge group is
\begin{equation}
G = SO(3) \times H \subset SO(3, n) \,,
\end{equation}
where the $SO(3)$ factor is related to the unbroken R-symmetry and \(H
\subset SO(n)\) has dimension \(\dim H \leq n\) and is specified by
\(f_{rst}\).\footnote{At the origin of the coset manifold
  \(\cM\) the coset representatives are simply delta-functions and the
  $SO(3)$ factor of the gauge groups corresponds indeed precisely to
  the $SU(2)$ R-symmetry. However generically \(L_I^i\) and \(L_I^r\)
  describe a non-trivial $SO(3,n)$ rotation and the $SO(3)$ factor
  does not need to coincide directly with the R-symmetry group, but
  can be  embedded into $SO(3,n)$ in a non trivial way.}
Since \(G\) is compact it automatically satisfies the condition
\eqref{eq:gaugecondition} and therefore is an allowed gauge group.

The generic case \(f_{ist}\neq 0\) is most conveniently analyzed if we go to a
specific basis for the vector multiplet index \(r\) where we can split \(r\) into
\(\hat r\) and \(\tilde r\) such that the only non-vanishing components of the
structure constants involving an \(\tilde r\) index are \(f_{\tilde r\tilde s \tilde t}\).
These components thus correspond to a group \(H \subset SO(q)\), \(q
\leq n\). The remaining components are \(f_{ijk}\), \(f_{i\hat r\hat
  s}\) and \(f_{\hat r\hat s\hat t}\) and they describe a non-compact group \(G_0 \subset SO(3,m)\), with \(m + q = n\) and \(SO(3) \subset G_0\).
The total gauge group 
then is 
\begin{equation}
G = G_0 \times H \subset SO(3, n) \,.
\end{equation}
If we furthermore assume that the gauge group is semi-simple, we can list all possible options for \(G_0\) explicitely. 
From \eqref{eq:gaugecondition} we know that \(G_0\) can have either at
most three compact or at most three non-compact generators and
the only non-compact semi-simple groups satisfying this condition and
containing \(SO(3)\) as a subgroup are \(SO(3,1)\) and \(SL(3,\bbR)\).  
Therefore $G$ has to be of the form
\begin{equation}\label{eq:gaugegroup}
G\ =\ G_0 \times H\ =\ \left\{ \begin{aligned} &SO(3) \\ &SO(3,1) \\ &SL(3, \bbR) \end{aligned}\right\} \times H\ \subset\ SO(3, n)\,,
\end{equation}
where \(H\) is an arbitrary semi-simple compact group.
This is in  agreement with the results from \cite{Karndumri:2014sha},
where however the compact factor \(H\) was not taken into account for
the analysis of AdS vacua.

\subsection{Moduli spaces of AdS backgrounds}\label{sugramoduli}

Let us now compute the moduli space 
of the AdS backgrounds determined
in the previous section. The moduli are the directions in the scalar manifold
$\cM$ given in 
\eqref{eq:cosetspace} which are undetermined by the conditions
\eqref{eq:Cconditions}. 
Or in other words we are looking for continuous solutions of the variations
\begin{equation}
\delta C_{ir} = 0\ ,\qquad \delta \left(e^{-\frac{5}{2}\sigma}
   C\right) = 0 \ ,\qquad \delta \xi^i=\delta \xi^r=0\ .
\end{equation}
The resulting scalar fields are automatically flat directions of the potential
\eqref{eq:potential} and thus can be viewed as the 
scalar degrees of freedom that remain massless in an AdS background.

We proceed along the lines of \cite{Louis:2014gxa} and parametrize the variations of the coset representatives as
\begin{equation}
\delta L_I^i = \left<L_I^r\right> \delta\phi_{ir} \,,
\end{equation}
where \(\delta\phi_{ir}\) are the fluctuations of the \(3n\) scalar fields around their background value.
Using \eqref{eq:eta} this implies
\begin{equation}
\delta L_I^r = \left<L_I^i\right> \delta\phi_{ir} \,,
\end{equation}
while the variations of the inverse coset representatives are given by
\begin{equation}
\delta L^I_i = -\left<L^I_r\right> \delta\phi_{ir} \,,\qquad \delta L^I_r = -\left<L^I_i\right> \delta\phi_{ir} \,.
\end{equation}

To simplify the notation we will from now on suppress the brackets and
assume that all field dependent  quantities are evaluated in the
background whenever this is appropriate. 
Since $\xi_I=0$ it follows directly that $\delta \xi^i=\delta \xi^r=0$
are satisfied without imposing any conditions on the variations of the
scalar fields.
From \eqref{abbreviations} and \eqref{eq:strconstcond}
we learn
\begin{equation}
\delta f_{ijk} = -3 f_{r[ij}\delta\phi_{k]r} = 0 \,,
\end{equation}
and thus \(\delta C=\delta\sigma = 0\).
The variation of \(C_{ir}\) on the other hand gives the non-trivial condition
\begin{equation}\label{eq:modulicondition}
0 = \delta f_{ijr} = - f_{ijk}\delta\phi_{kr} + 2 f_{rs[i}\delta\phi_{j]s} \,.
\end{equation}
It has been shown in the appendix of \cite{Louis:2014gxa} that all
solutions 
of this equation are of the form
\begin{equation}\label{eq:moduli}
\delta \phi_{ir} = f_{irs} \lambda^s \,,
\end{equation}
where \(\lambda^s\) are arbitrary real parameters.
Hence the number of independent moduli is given by the rank of the \((3n\times n)\)
matrix \(f_{ir\,s}\).\footnote{The notation should be understood in such a way that the pair of indices \(ir\) labels the rows of the matrix  \(f_{ir\,s}\) while \(s\) labels its columns.}
Adopting the notation of the previous section we should denote them by \(\lambda^{\hat s}\) and \eqref{eq:moduli} becomes \(\delta \phi_{i\hat r} = f_{i\hat r\hat s} \lambda^{\hat s}\), \(\delta \phi_{i\tilde r} = 0\).

The structure constants \(f_{i\hat r\hat s}\) precisely correspond to the non-compact generators of \(G_0\).
Since the maximally compact subgroup of \(G_0\) is in every case given by \(SO(3)\), we see that the scalar deformations span the coset manifold
\begin{equation}\label{eq:falsemodulispace}
{\cal M}_{\delta\phi} = \frac{G_0}{SO(3)} \,.
\end{equation}
Let us denote by \(\tilde G\) the maximal subgroup of \(SO(3,n)\) that
leaves the gauge group \(G\) and hence the structure constants
invariant.
Therefore, acting with \(\tilde G\) on a solution of
\eqref{eq:strconstcond} gives a 
rotated solution. 
It is therefore not unexpected that \({\cal M}_{\delta\phi}\) is of the form of an
orbit of \(\tilde G\) acting on the scalar manifold $\cM$ given in \eqref{eq:cosetspace}.

We will now argue that all scalars given in \eqref{eq:moduli} are in
fact Goldstone bosons eaten by massive vector fields and thus no
physical moduli.
For this purpose we evaluate the gauged Maurer-Cartan form
\eqref{eq:maurercartan} in the AdS background to find
\begin{equation}\label{eq:massterm}
P_{ir} = L_i^I \dd L_{rI} + f_{irs} A^s \,,
\end{equation}
where  \(A^s = L_I^s A^I\).
This expression appears quadratically in the Lagrangian
\eqref{eq:lagrangian} and thus gives a mass term for every vector field \(A^s\) in the preimage of the matrix \(f_{ir\,s}\).
Adopting again our previous notation, \eqref{eq:massterm} reads
\(P_{i\hat r} = L_i^I \dd L_{\hat rI} + f_{i\hat r\hat s} A^{\hat s}\), \(P_{i\tilde r} = L_i^I \dd L_{\tilde rI}\)
and we see that there is precisely one massive vector field \(A^{\hat s}\)
for every scalar \(\lambda^{\hat s}\). So no physical massless
direction is left and the moduli space %
can  only consist of isolated points.

We can also understand this result directly without analyzing the condition \eqref{eq:modulicondition}.
The vectors that obtain a mass in the AdS vacuum are in one-to-one
correspondence with the non-compact generators of the gauge group
\(G\). Therefore the mass term \eqref{eq:massterm} breaks the gauge group spontaneously to its maximally compact subgroup, i.e.
\begin{equation}
G = G_0 \times H \rightarrow SO(3) \times H \,.
\end{equation}
Breaking \(G_0\) to \(SO(3)\) in \eqref{eq:falsemodulispace} indeed
reduces 
${\cal M}_{\delta\phi}$ to a single point.


\section{The conformal manifold of the dual SCFT}\label{sec:SCFT}

In this section we study six-dimensional \(\cN = (1,0)\)
superconformal field theories (SCFTs)
which can serve as holographic duals
of the AdS backgrounds studied in the previous section.
In particular we focus on  possible marginal deformations of such
SCFTs which preserve 
all supercharges. We will however show that the representation theory of  the 
\(\cN = (1,0)\) superconformal algebra forbids any
such operators and thus no exactly marginal supersymmetric
deformations exist. This is equivalent to the statement 
that there is no conformal manifold ${\cal C}$. 
The AdS/CFT dictionary relates ${\cal C}$ to the 
moduli space 
of the dual AdS backgrounds which we studied
in the previous section.
As on both sides we only find vanishing deformation spaces our results show
perfect agreement.

\subsection{Preliminaries}\label{prelim}

Given a SCFT 
we can deform it by adding conformal operators \(\cO_i\) 
to the theory
\begin{equation}\label{eq:SCFTdeformations}
\cL \rightarrow \cL + \lambda^i \cO_i \ .
\end{equation}
$\cL$ denotes the Lagrangian but this notation is somewhat symbolic as we
  also consider SCFTs which do not have a Lagrangian description.
Operators \(\cO_i\) that do not break (super-) conformal invariance
are called exactly marginal operators. The space spanned by the
corresponding exactly marginal couplings \(\lambda^i\) is called the
conformal manifold \({\cal C}\).

A necessary condition for unbroken conformal invariance is that the
\(\lambda^i\) are dimensionless or equivalently that the operators
\(\cO_i\) 
have conformal dimension \(\Delta = 6\), i.e.\ are marginal operators.
This criterion is however not sufficient since higher-order
corrections in \(\lambda^i\) can perturb $\Delta$. 
In the following analysis we only consider marginal operators which do
not
break the \(\cN = (1,0)\) supersymmetry.
Thus the \(\cO_i\) of interest have to be the highest component of a
supermultiplet or in other words have to be annihilated by all
supercharges. 
In addition they should be singlets of the R-symmetry group.
The superconformal group  of six-dimensional \(\cN = (1,0)\) SCFTs
is the group  \(OSp(6,2|2)\) and its 
representations have been described in detail in
\cite{Minwalla:1997ka, Dobrev:2002dt, Bhattacharya:2008zy}. 
Let us briefly recall some of their results which we need 
for the following discussion.

The bosonic subalgebra of \(OSp(6,2|2)\) is \(SO(6,2)\times SU(2)_R\),
where \(SO(6,2)\) is the six-dimen\-sional conformal algebra and
\(SU(2)_R\) is the R-symmetry.
The fermionic part  of \(OSp(6,2|2)\) is generated by the supercharges
\( (Q^i_\alpha, S_i^\alpha)\) where $Q^i_\alpha$ is an R-doublet of chiral spinors with conformal
dimension \(\Delta=+ \frac{1}{2}\), while \( S_i^\alpha\) is an R-doublet of antichiral spinors with  \(\Delta=-\frac{1}{2}\).
Here \(\alpha = 1,\ldots,4,\) denotes
the fundamental representation of \(SU(4) = Spin(6)\) and \(i=1,2\) 
labels the fundamental representation of the \(SU(2)_R\).
The representation theory of the superconformal algebra is most
conveniently analyzed for the Euclidean theory, where one has the
Hermiticity relation \(Q^\dagger = S\), so that \(Q^i_\alpha\) and
\(S_i^\alpha\) can be interpreted as ladder operators, raising and
lowering the conformal dimension $\Delta$ by \(\frac{1}{2}\). 

As a consequence the unitary irreducible representations of
\(OSp(6,2|2)\)
decompose into direct sums of representations of
the maximally compact subgroup \(\SO(2) \times SO(6) \times \SU(2)_R\)
of the bosonic subgroup.  
Each representation can be built from a lowest weight state (conventionally called superconformal primary), which is characterized by the requirement that it is annihilated by all superconformal charges \(S_i^\alpha\).
Each primary is labeled by its conformal dimension \(\Delta_0\), three
half-integer \(SO(6)\) weights \(h_i=(h_1, h_2, h_3)\) 
 and a half-integer \(SU(2)\) weight \(k\).\footnote{It is sometimes convenient to translate the \(SO(6)\)  weights  \((h_i)\) into \(SU(4)\) Dynkin labels \([a_1 a_2 a_3]\) via
$
a_1 = h_2 - h_3 \,,\
a_2 = h_1 + h_2 \,,\
a_3 = h_2 + h_3 \,.
$
This implies in particular that they are not completely arbitrary but that they need to satisfy the constraint \(h_1 \geq h_2 \geq \left|h_3\right|\).
For example \((\frac{1}{2}, \frac{1}{2}, \pm\frac{1}{2})\) denotes the
(anti-)chiral spinor representation, while \((1,0,0)\) is the
\(SO(6)\) vector representation.}
The corresponding supermultiplet is then obtained 
by successively acting  with the supercharges \(Q^i_\alpha\)  on a superconformal primary.
A state obtained by the action of \(l\) supercharges is called a
level-\(l\) descendant and it has conformal dimension \(\Delta = \Delta_0 + \frac{l}{2}\).

Notice that \(\Delta_0\), \(h_i\) and \(k\) can be 
used to label the entire supermultiplet. 
It is however not possible to pick arbitrary combinations of values since unitarity imposes certain constraints.
Using the superconformal algebra 
(see Appendix~\ref{app:superconfalgebra})
one can compute the norm of the descendant states.
Requiring then that all states in a given representation have
non-negative norm implies bounds on the conformal dimension
\(\Delta_0\) of the primary operators which have the generic form
\begin{equation}\label{bound}
\Delta_0 \geq f(h_i, k) \,.
\end{equation}
The function \(f\) is explicitly determined in \cite{Minwalla:1997ka,
  Dobrev:2002dt} and we recall the results relevant for our analysis in the following section.
Representations that saturate the bound \eqref{bound} are short, as in this case some states have vanishing norm and are no longer part of the irreducible representation.
 
\subsection{Classification of marginal operators}\label{sec:marginaloperators}

After these preliminaries let us go in detail through all possible
candidates for supersymmetric marginal operators.
As we discussed, they must be part of a unitary representation of the 
superconformal algebra and therefore  are either  primary operators or
descendant operators
that are obtained by acting with \(l\) supercharges \(Q^i_{\alpha}\) on
a primary operator. 
However, the primary operators that are invariant under Lorentz-symmetry, R-symmetry and
supersymmetry have been shown to be proportional to the identity
operator \cite{Green:2010da}. 
Therefore we can restrict our further analysis to descendant operators.
Among the descendant operators we should also discard those operators
where two of the supercharges can be traded for a momentum operator by
means of the supersymmetry algebra.
These operators add in \eqref{eq:SCFTdeformations} only a total derivative
to the Lagrangian and hence do not deform the theory.
For the same reason the order of supercharges in a descendant operator
does not matter for our analysis.

If we start with a primary operator with \(SO(6)\) weights
\((h_1,h_2,h_3)\) we can only find Lorentz invariant descendant
operators at level
\begin{equation}\label{eq:invlevel}
l=2(h_1+h_2+h_3) +4n \,,
\end{equation}
with $n$ being an arbitrary non-negative integer.
In Appendix~\ref{app:minlevel} we give a proof of this statement.
Thus the conformal dimension of the primary operator needs to be
\begin{equation}\label{eq:confdimdesc}
\Delta_0 = 6 - \frac{l}{2} = 6 - h_1 - h_2 - h_3 - 2 n \,. 
\end{equation}
Moreover, we will use in the following that
\(k=0\) is only possible if \(l\) is even as 
descendants with an odd number of supercharges cannot be R-singlets.
The general bound from \cite{Minwalla:1997ka, Dobrev:2002dt} for a unitary representation reads 
\begin{equation}\label{eq:generalbound}
\Delta_0 \geq h_1 + h_2 - h_3 + 4k + 6\,,
\end{equation}
which is not compatible with \eqref{eq:confdimdesc}, since \(h_1\) and
\(h_2\) are necessarily non-negative. Therefore all descendants
of primary operators in long representations are excluded.

For special choices of the weights \((h_1, h_2, h_3)\) there
exist isolated short representations which we now turn to.
The following cases can be distinguished.

a) If \(h_1 - h_2 > 0\) and \(h_2 = h_3\), there is a short representation with
\begin{equation}
\Delta_0 = h_1 + 4k + 4 \,.
\end{equation}
Together with \eqref{eq:confdimdesc} the only possible solution is
\begin{equation}
(h_1, h_2, h_3) = (1, 0, 0)\,,\quad k=0\,,\quad\Delta_0 = 5 \,.
\end{equation}
A primary operator with these properties carries no R-symmetry indices and 
has to be an antisymmetric $SU(4)$-tensor (which is isomorphic to
the six-dimensional vector representation of $SO(6)$).
Thus the corresponding candidate descendant operator has to take the form
\begin{equation}
\cO_2 = \epsilon^{\alpha\beta\gamma\delta}\left\{Q_{i\alpha},[Q^i_\beta,U_{[\gamma\delta]}]\right\} \,,
\end{equation}
where $U_{[\gamma\delta]}$ is the associated primary operator with $\Delta_0=5$.
The norm of this operator can be computed straightforwardly
by using 
the superconformal algebra given in
Appendix~\ref{app:superconfalgebra}
with the result \(\left\|\cO_2\right\| \sim \Delta_0-5=0\).
As zero-norm states are not allowed in a unitary theory,
the operator \(\cO_2\) has to vanish.\footnote{Note
that this operators is a total derivative for any $\Delta_0$.
This is the case because the contraction of the R-symmetry indices is performed with an $\epsilon$-symbol, so \(\cO_2\) is symmetric under the exchange of the two supercharges and using \eqref{eq:qanticom} we see that \(\cO_2 \sim \left[P^{\alpha\beta}, U_{\alpha\beta}\right]\).}

b) For \(h_1 = h_2 = h_3 = h \neq 0\) there are additional short representations if
\begin{subequations}
\begin{align}
\Delta_0 &= 2 + h + 4k\ ,\qquad \text{or} \label{eq:boundc1}\\
\Delta_0 &= 4 + h + 4k\ . \label{eq:boundc2}
\end{align}
\end{subequations}
While \eqref{eq:boundc2} is not compatible with \eqref{eq:confdimdesc}, there are two solutions for \eqref{eq:boundc1}, namely 
\begin{equation}\label{sol1}
h=\frac{1}{2}\,,\quad k=\frac{1}{2}\,,\quad \Delta_0 = \frac{9}{2} \,,
\end{equation} 
and
\begin{equation}\label{sol2}
h=1\,,\quad k=0\,,\quad \Delta_0 = 3 \,.
\end{equation}
Denoting the primary operator for the first solution \eqref{sol1}
 by \(U^i_\alpha\), 
it is indeed possible to identify a Lorentz and R-symmetry invariant
descendant operator \(\cO_3\) at level \(l=3\) 
\begin{equation}
\cO_3 = \epsilon^{\alpha\beta\gamma\delta}\left\{Q_{i\alpha},\left[Q^i_\beta,\{Q_{j\gamma}, U^j_\delta\}\right]\right\} \,.
\end{equation}
Computing the norm yields
\(
\left\|\cO_3\right\| \sim \left(\Delta_0 - \frac{9}{2}\right)\left(\Delta_0 + \frac{7}{2}\right)
\)
and hence vanishes at the critical value \(\Delta_0 = 6-\frac{l}{2}
=\frac{9}{2}\). Consequently \(\cO_3\) itself vanishes and cannot be
considered as a possible marginal operator.
Notice that it is in principle possible to contract the R-symmetry
indices in a different fashion but all such operators 
differ from \(\cO_3\) only by a total derivative.
Moreover, we have checked that all these other combinations also have
vanishing norm. 

For the second solution \eqref{sol2} the primary operator has the
index structure \(U_{(\alpha\beta)}\) (with \(h=1\) and \(k=0\)) and we can build a Lorentz and R-symmetry invariant descendant operator \(\cO_6\) at level \(l=6\),
\begin{equation}
\cO_6 = \epsilon^{\alpha\beta\gamma\delta} \epsilon^{\epsilon\zeta\eta\theta} \left\{Q_{i\alpha},\left[Q^i_\epsilon,\left\{Q_{j\beta},\left[Q^j_\zeta,\left\{Q_{k\gamma},[Q^k_\eta, U_{(\delta\theta)}]\right\}\right]\right\}\right]\right\} \,.
\end{equation}
There are also other possibilities to contract the indices within \(\cO_6\), which would however lead to total derivatives.
In any case all these \(l=6\) operators are descendants of the operator \([Q_{i[\alpha},U_{(\beta]\gamma)}]\), whose norm is \((\Delta_0 - 3)\) and hence vanishes.

c)
Finally for \(h_1 = h_2 = h_3 = 0\) there are short representations for
\begin{equation}\label{noname}
\Delta_0 = 4k\,,\qquad \Delta_0 = 4k+2\,,\qquad \Delta_0 = 4k+4 \,.
\end{equation}
Since we have eight distinct supercharges, a descendant operator at
level \(l > 8\) is always zero by means of \eqref{eq:qanticom},
so according to \eqref{eq:invlevel} the only levels at
which we should look for suitable candidate operators are \(l = 4,8\).

At level \(l = 4\) we need $\Delta_0=4$ and there is one operator with \(k = 0\),
\begin{equation}\label{eq:O4}
\cO_4 = \epsilon^{\alpha\beta\gamma\delta} \left\{Q_{i\alpha},\left[Q^i_\beta,\left\{Q_{j\gamma},\left[Q^j_\delta,U\right]\right\}\right]\right\} \,,
\end{equation}
which has norm \(\left\|\cO_4\right\| \sim \Delta_0(\Delta_0-2)\). 
It does not vanish for \(\Delta_0 = 4\), but we find that the norm of \(\left[Q^i_\alpha, \cO_4\right]\) is proportional to \(\Delta_0(\Delta_0 - 2)(\Delta_0 + 1)\), so \(\left[Q^i_\alpha, \cO_4\right]\) vanishes only if \(\cO_4\) itself vanishes.
This means that \(\cO_4\)  breaks supersymmetry and thus cannot be a
supersymmetric marginal operator. 
Moreover \(\cO_4\) is also a total derivative. 

The only possibility for non-vanishing \(k\) is \(k=1\) as
\eqref{noname}
implies for $k>1$ that $\Delta_0>4$ while for $k=\frac12$ the level
$l$ cannot be even. 
The  operator with $k=1$ reads
\begin{equation}
\cO'_4 = \epsilon^{\alpha\beta\gamma\delta} \left\{Q_{i\alpha},\left[Q^i_\beta,\left\{Q_{j\gamma},\left[Q_{k\delta},U^{(jk)}\right]\right\}\right]\right\} \,.
\end{equation}
We can compute \(\left\|\cO'_4\right\| \sim (\Delta_0-4)(\Delta_0+6)(\Delta_0+8)\), and thus this operator is ruled out as well.
Clearly it is again also a total derivative.

At level \(l=8\) we need $\Delta_0=2$. Using the same argument as
above there is no operator with
\(k \neq 0\).
Hence a Lorentz invariant level \(l=8\) operator is (up to total
derivatives) always a descendant of the $l=2$ operator 
\begin{equation}
\cO^{ij}_{\alpha\beta} = \left\{Q^i_{[\alpha},\left[Q^j_{\beta]},U\right]\right\} \,.
\end{equation}
If we antisymmetrize also in the R-symmetry indices \(i\) and \(j\),
we find  \(\left\|\cO^{[ij]}_{\alpha\beta}\right\| \sim \Delta_0 \),
but this operator is symmetric under the exchange of the two supercharges
and we end up with a total derivative.
On the other hand we find for the symmetric component that
\(\left\|\cO^{(ij)}_{\alpha\beta}\right\| \sim \Delta_0 (\Delta_0 - 2)\),
so it vanishes at the dimension we are interested in.
Let us show for the sake of completeness that also all the level \(l=8\)
descendants of \(\cO^{[ij]}_{\alpha\beta}\) have vanishing or negative
norm at \(\Delta_0=2\).
They are in turn descendants of the \(l=4\) operator
\begin{equation}
\cO^{ij} = \epsilon^{\alpha\beta\gamma\delta} \left\{Q^i_\alpha,\left[Q^j_\beta, \epsilon_{kl}\cO^{kl}_{\gamma\delta}\right]\right\}
= \epsilon^{\alpha\beta\gamma\delta} \left\{Q^i_\alpha,\left[Q^j_\beta,\left\{Q_{k\gamma},\left[Q^k_\delta,U\right]\right\}\right]\right\} \,.
\end{equation}
While the antisymmetric part \(\cO^{[ij]}\) of this operator is nothing else
than \(\cO_4\) from \eqref{eq:O4} with norm \(\Delta_0(\Delta_0-2)\),
the symmetric part \(\cO^{(ij)}\) has norm \(\Delta_0(\Delta_0-2)(\Delta_0-4)\),
and so both operators have vanishing norm for $\Delta_0=2$.

To conclude we have thus shown that all candidates for marginal
operators either have zero norm or are not supersymmetric.
Notice that most of the operators are also total derivatives but we
did not have to use this fact in our argument. 
Let us close with the observation that 
the above analysis can be easily extended to relevant operators with conformal dimension \(\Delta < 6\).
In this case the dimension of the primary operator needs to satisfy
\begin{equation}
\Delta_0 = \Delta - \frac{l}{2} < 6 - h_1 - h_2 - h_3 - 2 n \,,\qquad n \in \bbN \,,
\end{equation}
which is clearly also not compatible with the general bound \eqref{eq:generalbound}.
Moreover for generic \(\Delta < 6\) all isolated short representations are ruled out as well.
Only for \(\Delta = 4\) the operators from c) with \(k=0\) remain possible candidate operators,
but we have shown that their norms are negative at the appropriate dimensions.

\section*{Acknowledgments}
This work was supported by the German Science Foundation (DFG) under
the Collaborative Research Center (SFB) 676 ``Particles, Strings and the Early
Universe'' and the Research Training Group (RTG) 1670 ``Mathematics
inspired by String Theory and Quantum Field Theory''.

We have benefited from conversations and correspondence with Ofer Aharony,
Gleb Arutyunov, Vincente Cort\'es, Ken Intriligator, Hans Jockers,
Constantin Muranaka, Christoph Schweigert, Alessandro Tomasiello
and especially Hagen Triendl and Marco Zagermann for pointing out an error in a previous version of this paper.

\newpage

\appendix
\noindent
{\bf\Huge Appendix}

\section{The $\cN = (1,0)$ superconformal algebra}
\label{app:superconfalgebra}

In this appendix we review the relevant (anti-) commutator relations
of the six-dimensional \(\cN = (1,0)\) superconformal algebra
\(OSp(6,2|2)\).
The conformal group \(SO(6,2)\) is generated by the Lorentz generators \(M_{\mu\nu}\), the momenta \(P_\mu\), the special conformal generators \(K_\mu\) and the dilatation operator \(D\).
The generators of the R-symmetry group \(SU(2)\) are denoted by \(R_i^j\),
where \(i,j = 1,2\).
In addition there are the supercharges \(Q^i_\alpha\), with \(\alpha = 1,\dots,4\), and the superconformal charges \(S_i^\alpha\), which together span the fermionic part of \(OSp(6,2|2)\).

It is convenient to use the local isomorphism \(SO(6) \cong SU(4)\) to label also the generators of the conformal group in an \(SU(4)\) covariant way,
i.e. the Lorentz generators become \(M^\alpha_\beta\) (with
$M^\alpha_\alpha=0$) and the momenta and special conformal generators become \(P_{[\alpha\beta]}\) and \(K_{[\alpha\beta]}\) respectively.

Since the commutation relations involving only bosonic operators are not relevant for our analysis and can be found for example in \cite{Minwalla:1997ka},
we only give the fermionic (anti-)commutators.
These are
\begin{equation}\begin{aligned}
\left[D, Q^i_\alpha\right] &= -\tfrac{i}{2}Q^i_\alpha \,, \\
\left[D, S_i^\alpha\right] &= \tfrac{i}{2}S_i^\alpha \,, \\
\left[M^\alpha_\beta, Q^i_\gamma\right] &= -i\left(\delta^\alpha_\gamma Q^i_\beta -\tfrac{1}{4}\delta^\alpha_\beta Q^i_\gamma\right)\,, \\
\left[M^\alpha_\beta, S_i^\gamma\right] &= i\left(\delta^\gamma_\beta S_i^\alpha -\tfrac{1}{4}\delta^\alpha_\beta S_i^\gamma\right) \,, \\
\left[R^i_j, Q^k_\alpha\right] &= -i\left(\delta^k_j Q^i_\alpha - \tfrac{1}{2}\delta^i_j Q^k_\alpha\right) \,, \\
\left[R^i_j, S_k^\alpha\right] &= i\left(\delta^i_k S_j^\alpha - \tfrac{1}{2}\delta^i_j S_k^\alpha\right) \,,
\end{aligned}\end{equation}
and
\begin{subequations}\begin{align}
\left\{Q^i_\alpha, Q^j_\beta\right\} &= \epsilon^{ij} P_{\alpha\beta} \,, \label{eq:qanticom}\\
\left\{S_i^\alpha, S_j^\beta\right\} &= \epsilon_{ij} K^{\alpha\beta} \,, \\
\left\{S_i^\alpha, Q^j_\beta\right\} &= i \left(2\delta_i^j M^\alpha_\beta - 4 \delta^\alpha_\beta R_i^j + \delta^\alpha_\beta \delta_i^j D \right) \,. 
\end{align}\end{subequations}

\section{Level of Lorentz-invariant descendant states}
\label{app:minlevel}

In this appendix we discuss at which levels it is possible to find a Lorentz-invariant descendant state, starting from a superconformal primary with given \(SO(6)\) weights \((h_1, h_2, h_3)\).
Let us denote the minimal level at which this is possible by \(N\) and notice that we will then also find Lorentz invariant states at the levels \(l = N + 4m\), \(m \in \bbN\).

The problem is conveniently analyzed in the language of \(SU(4)\)
Young tableau, since here \(N\) corresponds to the number of boxes
that need to be added to the diagram to fill up every of its columns
to the maximal length four. 
More generally if we switch to an arbitrary \(SU(n)\) Young tableau
and call the length of its \(i^{\rm th}\) row \(r_i\) and the length of its
\(i^{\rm th}\) column \(l_i\), \(N\) is given by 
\begin{equation}
N = \sum_{i=1}^{r_1} \left(n - l_i\right) \,,
\end{equation}
where the sum runs over all columns.
If we use the fact that the lengths of the columns and rows are related via
\begin{equation}
l_i = p \qquad {\rm for}\quad r_{p+1} < i \le r_p\ , \quad p=
1,\ldots,n-1\ ,
\end{equation}
and that the Dynkin labels \(a_i\) can by read off from the tableau by
\begin{equation}
a_i = r_i - r_{i+1} \,,
\end{equation}
where \(r_n \equiv 0\), we find
\begin{equation}
N = \sum_{i=1}^{n-1}\left(n-i\right)a_i \,.
\end{equation}
Going back to the relevant case \(n=4\) and using that \(a_1 = h_2 - h_3\), \(a_2 = h_1 + h_2\), \(a_3 = h_2 + h_3\), the result reduces to
\begin{equation}
N = 2 \left( h_1 + h_2 + h_3 \right) \,.
\end{equation}


\providecommand{\href}[2]{#2}\begingroup\raggedright\endgroup

\end{document}